\documentclass[aps,prx,twocolumn,superscriptaddress,showpacs]{revtex4-1}

\usepackage{color}

\usepackage{graphicx}

\begin{document}

\title{Method for Estimating Spin-Spin Interactions from Magnetization Curves}

\author{Ryo Tamura}
\email[]{tamura.ryo@nims.go.jp}
\affiliation{Computational Materials Science Unit, National Institute for Materials Science, 1-1 Namiki, Tsukuba, Ibaraki 305-0044, Japan}
\affiliation{Center for Materials Research by Information Integration,
National Institute for Materials Science, 1-2-1 Sengen, Tsukuba, Ibaraki 305-0047, Japan}

\author{Koji Hukushima}
\email[]{hukusima@phys.c.u-tokyo.ac.jp}
\affiliation{Department of Basic Science, Graduate School of Arts and Sciences, The University of Tokyo, Komaba, Meguro, Tokyo 153-8902, Japan}
\affiliation{Center for Materials Research by Information Integration,
National Institute for Materials Science, 1-2-1 Sengen, Tsukuba, Ibaraki 305-0047, Japan}

\date{\today}

\begin{abstract}
We develop a method to estimate the spin-spin interactions in the Hamiltonian from the observed magnetization curve by machine learning based on Bayesian inference. 
In our method, plausible spin-spin interactions are determined by maximizing the posterior distribution, which is the conditional probability of the spin-spin interactions in the Hamiltonian for a given magnetization curve with observation noise.
The conditional probability is obtained by the Markov-chain Monte Carlo simulations combined with an exchange Monte Carlo method. 
The efficiency of our method is tested using synthetic magnetization curve data, 
and the results show that spin-spin interactions are estimated with a high accuracy.
In particular, 
the relevant terms of the spin-spin interactions are successfully selected from the redundant interaction candidates by the $l_1$ regularization in the prior distribution. 
\end{abstract}



\maketitle

\section{Introduction}

\begin{figure}[b]
\begin{center}
\includegraphics[scale=1.0]{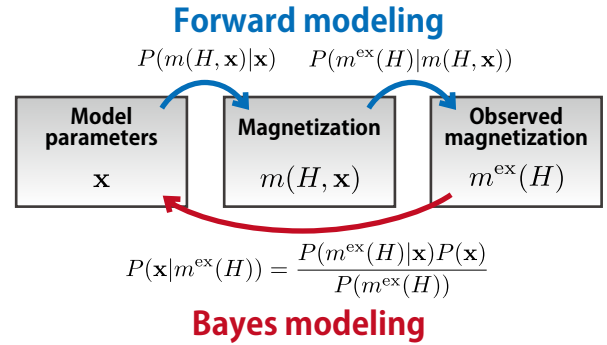} 
\end{center}
\caption{\label{fig:forward}
(Color online)
Schematic research flows for forward modeling and Bayes modeling.
$P(A)$ is the probability distribution of $A$, and $P(A|B)$ is the conditional probability of $A$ given $B$. 
}
\end{figure}

The importance of data-driven techniques using machine learning\cite{Mitchell-1997,Bishop-2006} is recognized in both academic and industrial fields.
Machine learning is generally a statistical tool used to extract the inherent structure from a finite set of observed data. 
In the condensed matter physics,
machine learning techniques have been used such as for interpolation of
density-functional theory
(DFT)\cite{Snyder-2012,Bartok-2013,Zhou-2014,Li-2015} and dynamical-mean
field theory\cite{Arsenault-2014} calculations, and for model selection of strongly correlated systems\cite{Takenaka-2014} and the Ginzburg-Landau equation\cite{Taverniers-2015}.
In this paper,
we apply Bayesian inference from the field of machine learning \cite{Toussaint-2011,DAgostini-2003,Dose-2003} to estimate the spin-spin interactions in the Hamiltonian given an observed magnetization curve as an input (Fig.~\ref{fig:forward}).

Magnetization is the average of the magnetic moments normalized by the amplitude of the magnetic moment, 
which is induced in a magnetic material by a magnetic field and is a fundamental physical quantity in both experiments and theories of magnetism. 
The magnetic-field dependence of the magnetization is called the magnetization curve\cite{Nagamiya-1962,Kanamori-1966}, 
and has been measured in many magnetic materials using devices such as a superconducting quantum interference device (SQUID) in the laboratory.
In general, the magnetization curve for a fixed temperature increases monotonically as a nonlinear function of the magnetic field.
Depending on the magnetic material, various characteristic features, 
such as plateaus\cite{Hida-1994,Oshikawa-1997,Totsuka-1998,Kitazawa-1999,Onizuka-2000,Kikuchi-2005,Kikuchi-2005,Ishii-2011}, 
ramps\cite{Bindilatti-1996,Gratens-2004,Nakano-2010}, 
and sharp increases\cite{Katsumata-1989},
may be observed in the magnetization curve. 
Consequently, the magnetization curve is an important quantity for understanding the magnetic properties of individual magnetic materials.

%

To understand the essence of the properties in magnetic materials, an effective Hamiltonian is often derived in materials science.
Many methods have been proposed to determine Hamiltonians, and they can be divided roughly into two groups:
one where the model parameter values in the Hamiltonian are calculated by ab initio electronic structure calculations when basic data of the magnetic materials is inputted\cite{Mazin-2007,Nakamura-2008,Pruneda-2008,Hirayama-2013}, 
and the other where the model parameter values are evaluated by fitting the physical quantities observed in a target magnetic material\cite{Takubo-2007,Tamura-2011,Matsumoto-2012,DiStasio-2013,Marcotte-2013}.

The aim of this paper is to establish a new Hamiltonian estimation method, which can be classified as the latter group where the model parameters in the Hamiltonian of a target material are predicted from experimental data  (Fig.~\ref{fig:forward}).
In particular,
we propose a method, which selects relevant terms for the model parameters from redundant model parameter candidates by machine learning based on the $l_1$ regularization.
The framework of this method can be adopted for any measurement data,
but herein we focus on the case where an observed magnetization curve is used as the input data to estimate the spin-spin interactions.

The rest of the paper is organized as follows. 
In Sec.~II,
we introduce a method to estimate the model parameters in the Hamiltonian from the observed magnetization curve.
We first present the forward modeling for an observed magnetization curve.
By using Bayesian inference,
we construct the posterior distribution, 
that is, the conditional probability of the model parameters in the Hamiltonian given an observed magnetization curve (the Bayes modeling).
In this case,
the plausible model parameters maximize the posterior distribution.
Next, we explain the simulation methods, which employ the Markov-chain Monte Carlo (MCMC) method and exchange Monte Carlo method, to analyze the posterior distribution.
In the last part of Sec.~II, we describe cross validation to avoid over-fitting.
Cross validation makes it possible to calculate prediction errors and allows a hyperparameter in the prior distribution to be determined.
To validate our estimation method,
we determine the spin-spin interactions from a synthetic magnetization curve in Sec.~III.
We prepare an input observed magnetization curve using a theoretical model with fixed spin-spin interactions to consider the case where the solution of the estimation problem is known in advance.
We show that the spin-spin interactions can be accurately estimated from the observed magnetization curve using our method.
Section V presents the discussion and summary.

\section{Method for estimating model parameters}

This section presents our estimation method for model parameters in the Hamiltonian determined from the observed magnetization curve.
Using Bayesian inference, the conditional probability of the spin-spin interactions are obtained for a given observed magnetization curve.
Furthermore, the simulation methods used to analyze the conditional probability and the cross validation to avoid over-fitting are introduced.

\subsection{Bayesian inference for the magnetization curve}

\subsubsection{Forward modeling}
Let us assume a class of the Hamiltonian $\mathcal{H}(\{\mathbf{s}_i\};\mathbf{x})$ for a magnetic system with a set of model parameters denoted as $\mathbf{x}=(x_1, ..., x_K)$ where $K$ is the number of model parameters.
Here,
$\mathbf{s}_i$ is the spin variable at site $i$ with the absolute value $|\mathbf{s}|$, and $\{\mathbf{s}_i\}$ denotes a set of spin configurations.
In statistical physics,
the magnetization under a magnetic field $H$ at a temperature $T$ for a given Hamiltonian $\mathcal{H} (\{\mathbf{s}_i\};\mathbf{x})$ is calculated as
\begin{eqnarray}
&&m (H,\mathbf{x}) = \left| \frac{1}{N |\mathbf{s}|} \sum_{i=1}^N \langle \mathbf{s}_i \rangle_{H,\mathbf{x}} \right|, \\
&&\langle \mathbf{s}_i \rangle_{H,\mathbf{x}} = \frac{\displaystyle{{\rm Tr} \ \mathbf{s}_i \exp \left[- \left( \mathcal{H} (\{\mathbf{s}_i\};\mathbf{x}) - H \sum_{i=1}^N s_i^z \right)/T \right]}}{\displaystyle{{\rm Tr}  \exp \left[-\left( \mathcal{H} (\{\mathbf{s}_i\};\mathbf{x}) - H \sum_{i=1}^N s_i^z \right)/T \right]}}, \ \ \ \ 
\end{eqnarray}
where $s_i^z$ is the $z$-component of $\mathbf{s}_i$ and $N$ is the number of spins. 
Throughout the paper, 
the $g$-factor, the Bohr magneton, and the Boltzmann constant are set to
unity.
In the forward modeling,
the conditional probability of the observed magnetization in the given model parameters is examined. 
Because the magnetization $m (H,\mathbf{x})$ is uniquely obtained in the thermodynamic limit by statistical physics,
the conditional probability of the magnetization $m (H,\mathbf{x})$ given $\mathbf{x}$ is expressed as
\begin{eqnarray}
P (m (H,\mathbf{x}) | \mathbf{x}) = \delta \left( m (H,\mathbf{x}) - \left| \frac{1}{N |\mathbf{s}|} \sum_{i=1}^N \langle \mathbf{s}_i \rangle_{H,\mathbf{x}} \right| \right), \ \ 
\end{eqnarray}
where $\delta (\cdots)$ is the Dirac delta function and $P(A|B)$ expresses the conditional probability of $A$ given $B$. 

The measurements in an experiment always have some uncertainty.
Taking this uncertainty into account as the observation noise $\epsilon$,
the observed magnetization in experiments $m^{\rm ex} (H)$ is expressed as 
\begin{eqnarray}
m^{\rm ex} (H) = m (H,\mathbf{x}) + \epsilon.
\end{eqnarray}
At this point, we assume that the observation noise follows a Gaussian distribution with a mean of zero and a standard deviation of $\sigma$:
\begin{eqnarray}
P (\epsilon) \propto \exp \left( - \frac{\epsilon^2}{2 \sigma^2} \right).
\end{eqnarray}
The conditional probability of $m^{\rm ex} (H)$ for a given $m (H,\mathbf{x})$ is then
\begin{eqnarray}
&&P (m^{\rm ex} (H) | m (H,\mathbf{x})) \nonumber \\
&&\ \ \ \ \ \ \ \ \ \ \ \ \ \ \ \propto \exp \left[ - \frac{1}{2 \sigma^2} \left( m^{\rm ex} (H) - m (H,\mathbf{x}) \right)^2 \right].
\end{eqnarray}
Thus, in the forward modeling, 
the conditional probability of the observed magnetization $m^{\rm ex} (H)$ given the model parameters $\mathbf{x}$ is written as
\begin{eqnarray}
&&P (m^{\rm ex} (H) | \mathbf{x}) \nonumber \\
&& \ \ \ \ \  \propto \int d m (H,\mathbf{x}) P (m^{\rm ex} (H) | m (H,\mathbf{x})) P (m (H,\mathbf{x}) | \mathbf{x}) \\
&& \ \ \ \ \  \propto \exp \left[ - \frac{1}{2 \sigma^2} \left( m^{\rm ex} (H) - \left| \frac{1}{N |\mathbf{s}|} \sum_{i=1}^N \langle \mathbf{s}_i \rangle_{H,\mathbf{x}} \right| \right)^2 \right]. \ \ \ \ \
\end{eqnarray}
By using this formula,
the probabilistic prediction of the magnetization observed in the experiments is obtained as the conditional probability for the given model parameters.
This can be compared with the experimental data.
Figure~\ref{fig:forward} shows the flow of the forward modeling.

\subsubsection{Bayes modeling}

The Bayes modeling provides a general framework to estimate the model parameters $\mathbf{x}$ from the observed magnetizations.
In this work, we consider the case where the magnetization under various magnetic fields is observed experimentally;
that is,
the observed magnetization curve is expressed as $\{ m^{\rm ex} (H_l) \}_{l\in D}$ for a series of magnetic fields $\{H_l\}_{l\in D}$ 
where the data are indexed by $D=(1, ..., L)$ and $L$ is the number of observed magnetizations.
By using Bayes' theorem,
the posterior distribution, which is the conditional probability of the model parameters $\mathbf{x}$ given the observed magnetization curve $\{ m^{\rm ex} (H_l) \}_{l\in D}$,  is expressed as
\begin{eqnarray}
P (\mathbf{x} | \{ m^{\rm ex} (H_l) \}_{l\in D}) = \frac{P(\{ m^{\rm ex} (H_l) \}_{l\in D} | \mathbf{x}) P (\mathbf{x})}{P(\{ m^{\rm ex} (H_l) \}_{l\in D})}, \ \
\end{eqnarray}
where $P(\mathbf{x})$ is the prior distribution of $\mathbf{x}$.
On the right-hand side, the denominator is independent of the model parameters. 
In contrast,
the prior distribution $P(\mathbf{x})$ explicitly contributes to the estimation in the posterior distribution. 
For example,
in the case without a priori knowledge of the model parameters,
a uniform distribution of $P (\mathbf{x})$ may be used.
In this work, 
we assume that the number of important model parameters is small, 
indicating that the relevant parameters are sparse in the model Hamiltonian.
In this situation, 
the prior distribution $P (\mathbf{x})$ based on the $l_1$ regularization is frequently represented as
\begin{eqnarray}
P (\mathbf{x}) \propto \exp \left( - \lambda \sum_{k=1}^K | x_k | \right), \label{eq:lasso}
\end{eqnarray}
where $\lambda$ is called the hyperparameter given before the analysis.
The inference scheme using the prior distribution is called LASSO (least absolute shrinkage and selection operator)\cite{Tibshirani-1996}, and the value of $\lambda$ controls the strength of the regularization.

In addition, 
we assume that $m^{\rm ex} (H_l)$ is independently obtained from different magnetic fields.
That is, the previous measurement in a magnetization process does not affect the present measurement,
which gives 
\begin{eqnarray}
P ( \{ m^{\rm ex} (H_l) \}_{l\in D} | \mathbf{x}) = \prod_{l=1}^L P ( m^{\rm ex} (H_l) | \mathbf{x}).
\end{eqnarray}
Combined with the prior distribution of Eq.~(\ref{eq:lasso}), 
the posterior distribution of $\mathbf{x}$ given $\{ m^{\rm ex} (H_l) \}_{l\in D}$ based on Bayes' theorem is expressed as 
\begin{eqnarray}
&&P (\mathbf{x} | \{ m^{\rm ex} (H_l) \}_{l\in D}) \nonumber \\
&&\ \propto \prod_{l=1}^L P ( m^{\rm ex} (H_l) | \mathbf{x}) P (\mathbf{x}) \\
&&\ \propto \exp \Biggl[ - \frac{1}{2 \sigma^2} \sum_{l=1}^L \left( m^{\rm ex} (H_l) - \left| \frac{1}{N |\mathbf{s}|} \sum_{i=1}^N \langle \mathbf{s}_i \rangle_{H_l,\mathbf{x}} \right| \right)^2 \nonumber \\
&&\ \ \ \ \ \ \ \ \ \ \ \ \ \ \ \ \ \ \ \ \ \ \ \ \ \ \ \ \ \ \ \ \ \ \ \ \ \ \ \ \ \ \ \ \ \ \ \ \ \ \ - \lambda \sum_{k=1}^K | x_k | \Biggr]. \ \ \ \ \ \ \label{eq:Pxm}
\end{eqnarray}
From the view point of the maximum a posterior (MAP) estimation,
the plausible model parameters $\mathbf{x}^*$ are obtained by the maximizer of Eq.~(\ref{eq:Pxm}).
In other words,
we search the model parameters so that Eq.~(\ref{eq:Pxm}) is maximized in the Bayes modeling where $\sigma$, $\lambda$, and $K$ are the previously fixed hyperparameters.
Although the posterior distribution is constructed using the magnetization curve as the input data,
it should be emphasized that the framework of our estimation method can be adopted for any measurement data.

\subsection{Simulation method}

In this study,
we use the MCMC method and the exchange Monte Carlo method to analyze the posterior distribution $P (\mathbf{x} | \{ m^{\rm ex} (H_l) \}_{l\in D})$.
This combination significantly contributes to finding the global maximum of the posterior distribution in systems where many local maxima exist.
When the energy function $E (\mathbf{x} | \sigma, \lambda, K)$ is defined as
\begin{eqnarray}
&&E (\mathbf{x} | \sigma, \lambda, K) \nonumber \\
&&\ \ \ \ := \frac{1}{2 \sigma^2} \sum_{l=1}^L \left( m^{\rm ex} (H_l) - \left| \frac{1}{N |\mathbf{s}|} \sum_{i=1}^N \langle \mathbf{s}_i \rangle_{H_l,\mathbf{x}} \right| \right)^2 \nonumber \\
&&\ \ \ \ \ \ \ \ \ \ \ \ \ \ \ \ \ \ \ \ \ \ \ \ \ \ \ \ \ \ \ \ \ \ \ \ \ \ \ \ \ \ \ \ \ \ + \lambda \sum_{k=1}^K | x_k |, \ \ 
\end{eqnarray}
the posterior distribution can be written as
\begin{eqnarray}
P (\mathbf{x} | \{ m^{\rm ex} (H_l) \}_{l\in D}) \propto \exp \left[ - \frac{1}{T_{\rm R}} E (\mathbf{x} | \sigma, \lambda, K) \right], \label{eq:Boltz}
\end{eqnarray}
where $T_{\rm R}=1$ in the case of Eq.~(\ref{eq:Pxm}).
This distribution formally has the same form as the Boltzmann distribution by regarding  $E (\mathbf{x} | \sigma, \lambda, K)$ and $T_{\rm R}$ as the energy of the system and the virtual temperature, respectively. 
Thus, a statistical mechanical approach is a promising for sampling from the distribution\cite{Landau-2009}.
It should be noted that the dynamical variables in this scheme are the model parameters in the Hamiltonian, 
which is unusual in the study of magnetism as well as in statistical physics.

In our MCMC simulations, 
we employ the Metropolis-type transition probability from $\mathbf{x}$ to $\mathbf{x}'$:
\begin{eqnarray}
&&w (\mathbf{x}'|\mathbf{x}) = \min \left\{1, \exp \left[- \frac{1}{T_{\rm R}} \Delta E(\mathbf{x}',\mathbf{x}) \right] \right\}, \label{eq:tran_p} \\
&&\Delta E(\mathbf{x}',\mathbf{x}) := E(\mathbf{x}' | \sigma, \lambda, K) - E (\mathbf{x} | \sigma, \lambda, K).
\end{eqnarray}
By sampling, 
we investigate the properties of $P (\mathbf{x} | \{ m^{\rm ex} (H_l) \}_{l\in D})$ and search for the model parameters $\mathbf{x}^*$ that maximize $P (\mathbf{x} | \{ m^{\rm ex} (H_l) \}_{l\in D})$.

In addition, 
we use the exchange Monte Carlo method\cite{Hukushima-1996} to enhance the sampling efficiency and to find the global maximum of $P (\mathbf{x} | \{ m^{\rm ex} (H_l) \}_{l\in D})$ in a system where many local maxima exist.
In the method, several MCMC simulations using Eq.~(\ref{eq:tran_p}) with different virtual temperatures are performed in parallel,
and the model parameters $\mathbf{x}$ are exchanged between the different MCMC simulations with an exchange probability expressed as
\begin{eqnarray}
&&w_{\rm ex} (\mathbf{x}^j|\mathbf{x}^i) = \min \left\{1, \ \exp \left[\left(\frac{1}{T_{\rm R}^j} - \frac{1}{T_{\rm R}^i} \right) \Delta E(\mathbf{x}^j,\mathbf{x}^i) \right] \right\}, \nonumber \\\label{eq:ex_p} \\
&&\Delta E(\mathbf{x}^j,\mathbf{x}^i) := E(\mathbf{x}^j | \sigma, \lambda, K) - E (\mathbf{x}^i | \sigma, \lambda, K),
\end{eqnarray}
where $\mathbf{x}^i$ and $T_{\rm R}^i$ indicate the model parameters and the virtual temperature in the $i$-th MCMC simulation, respectively.
By using this technique,
we can find the global maximum of $P (\mathbf{x} | \{ m^{\rm ex} (H_l) \}_{l\in D})$ and obtain the plausible model parameters $\mathbf{x}^*$ with a high probability.
Without a loss in generality, 
the units of the parameter $\lambda$ and the virtual temperature $T_{\rm R}$ can be taken as $\sigma^2$. 
Thus, we set $\sigma=1$ throughout the paper.

\subsection{Cross validation}

In general, 
the observed magnetization curve is well fitted when the number of model parameters is sufficiently large.
Often the estimated model in machine learning cannot properly predict unknown data not used in the fitting,
even when the observed data are well fitted by the estimated model. 
This is known as the over-fitting problem.

\begin{figure}[]
\begin{center}
\includegraphics[scale=1.0]{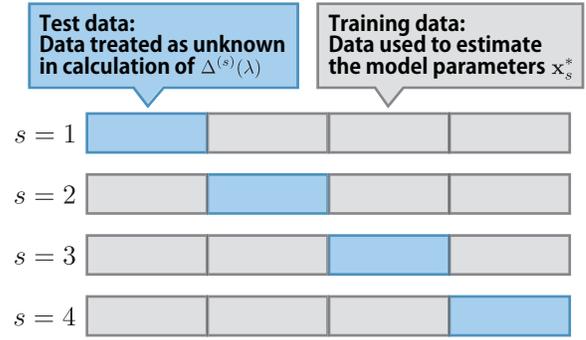} 
\end{center}
\caption{\label{fig:cross}
(Color online)
Cross validation for $S=4$.
We evaluate $\Delta^{(s)} (\lambda)$ in each case.
Average of $\Delta^{(s)} (\lambda)$ over the four cases is used to estimate $\lambda^*$.
}
\end{figure}

To determine the hyperparameter $\lambda$ while preventing over-fitting, 
we perform cross validation in which the hyperparameter is chosen to minimize the prediction error.
Consider the case where the value of $K$ is fixed.
In cross validation,
the dataset $D$ is divided into $S$ groups as
$D_s=((s-1)L/S+1, (s-1)L/S+2, ..., sL/S)$ with $s=1, ..., S$. 
One of the $S$ groups is regarded as the test data,
while the remaining $S-1$ groups are used as the training data. 
Figure~\ref{fig:cross} shows an example for $S=4$.
The number of pieces of test data and that of the training data are given by $L_{\rm te}:=N/S$ and $L_{\rm tr}:=N (S-1) /S$, respectively.
For each data subset $G_s=D\setminus D_s$,
the plausible model parameters $\mathbf{x}^*_s$ are determined by the MAP estimation according to $\lambda$.
The mean-square deviation between the test data and the estimated magnetization curve when using $\mathbf{x}^*_s$ for various $\lambda$ is evaluated as
\begin{eqnarray}
&&\Delta^{(s)} (\lambda) \nonumber \\
&&\ \ \ \ := \frac{1}{L_{\rm te}}
\sum_{l_{\rm te}\in D_s}
\left( m^{\rm ex} (H_{l_{\rm te}}) - \left| \frac{1}{N |\mathbf{s}|} \sum_{i=1}^N \langle \mathbf{s}_i \rangle_{H_{l_{\rm te}},\mathbf{x}^*_s} \right| \right)^2. \nonumber \\ 
\label{eq:delta_l}
\end{eqnarray}
In the cross validation, 
$\Delta^{(s)} (\lambda)$ represents the prediction error when the test data $D_s$ are treated as unknown data.
Using the average of $\Delta^{(s)}(\lambda)$ over different data sets, 
we evaluate the optimal value of $\lambda^*$ where the averaged prediction error takes a minimum as a function of $\lambda$. 
Furthermore, 
the plausible model parameters $\mathbf{x}^*$ are obtained such that $\Delta^{(s)} (\lambda)$ is the smallest among $S$ selections for the test data for $\lambda^*$.
It should be noted that $\lambda^*$ depends on the input magnetization curve by performing cross validation.

%

\section{Demonstration of estimation of spin-spin interactions}

In this section, 
we confirm that the estimation method proposed in Sec.~II can determine the spin-spin interactions in the Hamiltonian.
The inputted observed magnetization curve was a zero-temperature magnetization curve calculated from a theoretical model.
The results confirm that our estimation method is highly accurate and the relevant terms of the spin-spin interactions are successfully selected from the redundant interaction candidates by the $l_1$ regularization.

\subsection{Model and observed magnetization curve}

\begin{figure}[]
\begin{center}
\includegraphics[scale=1.0]{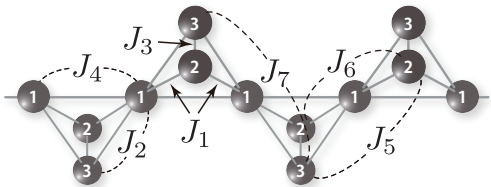} 
\end{center}
\caption{\label{fig:lattice}
(Color online)
Lattice structure and types of spin-spin interactions in the model Hamiltonian used to generate the synthetic magnetization curve.
Numbers indicate the types of lattice points.
}
\end{figure}

We examine the efficiency of our proposed method by synthetic data for the magnetization curve. 
To generate the input data,
we used the classical Heisenberg model with bilinear and biquadratic interactions,
which has the following Hamiltonian:
\begin{eqnarray}
\mathcal{H} = \sum_{i,j} J_{ij} \left[ \mathbf{s}_i \cdot \mathbf{s}_j - b_{ij} (\mathbf{s}_i \cdot \mathbf{s}_j)^2 \right] - H \sum_{i} s_i^z. \label{eq:ham}
\end{eqnarray}
Here, $b_{ij} = b J_{ij}$, and $\mathbf{s}_i = (s_i^x, s_i^y, s_i^z)$ denotes the three component vector spin with a length $|\mathbf{s}| =1/2$.
The value of $b$ represents the amplitude of the biquadratic interactions.
Depending on the lattice and sets of spin-spin interactions,
this model exhibits fruitful magnetization curves\cite{Penc-2004,Motome-2006,Shannon-2006}.
Here, the inputted observed magnetization curve is given by the model on a tetrahedral chain at a zero temperature.
The lattice structure is shown in Fig.~\ref{fig:lattice}.
For simplicity, we consider seven types of spin-spin interactions $J_k \ (k=1, ..., 7)$.
Thus, the  model parameter can be expressed by $\mathbf{x}=(J_1, J_2, J_3, J_4, J_5, J_6, J_7, b )$ with the dimension $K=8$.

\begin{figure}[]
\begin{center}
\includegraphics[scale=1.0]{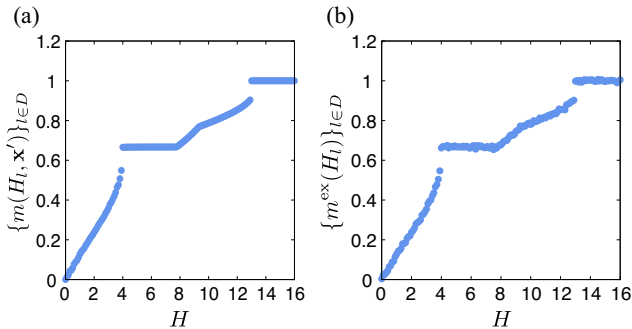} 
\end{center}
\caption{\label{fig:ob_mag}
(Color online)
(a) Magnetization curve $\{m(H_l, \mathbf{x}')\}_{l \in D}$ for $L=160$ at the zero temperature in the theoretical model defined by Eq.~(\ref{eq:ham}) with model parameter $\mathbf{x}'=(1, 4, 5, 6, 0, 0, 0, 0.1)$. 
(b) Observed magnetization curve $\{m^{\rm ex} (H_l)\}_{l \in D}$ represented by $\{m(H_l, \mathbf{x}')\}_{l \in D}$ with added Gaussian noise of the mean zero and standard deviation $0.004$.
}
\end{figure}

Figure~\ref{fig:ob_mag} (a) is the zero-temperature magnetization process $\{m(H_l, \mathbf{x}')\}_{l \in D}$ with the number of data points $L=160$,
which is obtained by
$\mathbf{x}'=(1, 4, 5, 6, 0, 0, 0, 0.1)$. 		     
The magnetization curve exhibits a huge 2/3 magnetization plateau.
Furthermore, we assume that the observed magnetization curve $\{m^{\rm ex} (H_l)\}_{l \in D}$ can be represented by adding Gaussian noise with a mean of zero and a standard deviation of $\sigma=0.004$. 
Figure~\ref{fig:ob_mag} (b) shows the synthesized magnetization curve $\{m^{\rm ex} (H_l)\}_{l \in D}$.


For the magnetization curve shown in Fig.~\ref{fig:ob_mag} (b), 
we estimate model parameters $\mathbf{x}=(J_1, J_2, J_3, J_4, J_5, J_6, J_7, b)$ using our proposed method.
For the cross validation,
we randomly divided the observed magnetization curve into four sets of data points $(S=4)$. 
Each set of training data contains 120 data points $(L_{\rm te}=120)$ in the magnetization curve.

\begin{table*}[]
\begin{center}
\caption{\label{tab:interaction}
Number of each type of interaction $J_k$ per tetrahedron, 
distance between spins connected by each interaction $J_k$ within the nearest-neighbor lattice spacing, 
spin-spin interactions used in the input magnetization curve, 
and spin-spin interactions estimated using our estimation method.
}
\begin{tabular}{l||c|c|c|c|c|c|c|c}
\hline\hline
&$\ \ \ \ J_1\ \ \ \ $& $\ \ \ \ J_2\ \ \ \ $ &$\ \ \ \ J_3\ \ \ \ $&$\ \ \ \ J_4\ \ \ \ $&$\ \ \ \ J_5\ \ \ \ $&$\ \ \ \ J_6\ \ \ \ $&$\ \ \ \ J_7\ \ \ \ $&$\ \ \ \ b\ \ \ \ $ \\
\hline\hline
Number of interactions $(n_k)$ &$2$&$2$&$1$&$1$&$2$&$1$&$1$&$-$ \\
\hline
Distance between spins $(r_k)$ &$1$&$1$&$1$&$1$&$\sqrt{3}$&$2$&$2$&$-$ \\
\hline\hline
Input spin-spin interactions ($\mathbf{x}'$) &$1.000$&$4.000$&$5.000$&$6.000$&$0.000$&$0.000$&$0.000$&$0.100$ \\
\hline
Estimated spin-spin interactions ($\mathbf{x}^*$) &$1.074$&$\ 3.850\ $&$\ 5.012\ $&$\ 6.051\ $&$\ 0.011\ $&$\ -0.051\ $&$\ 0.002\ $&$\ 0.102\ $ \\
\hline\hline
\end{tabular}
\end{center}
\end{table*}

\subsection{Estimation of spin-spin interactions}

This section explains the details of the numerical simulations to evaluate the posterior distribution $P (\mathbf{x} | \{ m^{\rm ex} (H_l) \}_{l\in G_s})$ with $s=1,...,4$.
Our numerical method consists of double loop calculations, 
which include a calculation of the magnetization curve for a given set of model parameters and the sampling of the model parameters. 
We used the Hamiltonian defined by Eq.~(\ref{eq:ham}) with periodic boundary conditions where the total number of spins was six as a theoretical model to obtain $\langle \mathbf{s}_i \rangle_{H_l,\mathbf{x}}$.
The zero-temperature magnetization curve was obtained by the steepest descent method.
This is the inner-loop calculation which was performed in a few seconds. 
Monte Carlo simulations were employed to sample the model parameters $\mathbf{x}$ in the outer loop.
Typically, 
the number of the Monte Carlo steps to update the model parameters was $10^{4}$.
Furthermore, 
we prepared $20$ replicas with virtual temperatures $T_{\rm R}$ between $0.001$ and $10$ for the exchange Monte-Carlo method.

\begin{figure*}[]
\begin{center}
\includegraphics[scale=1.0]{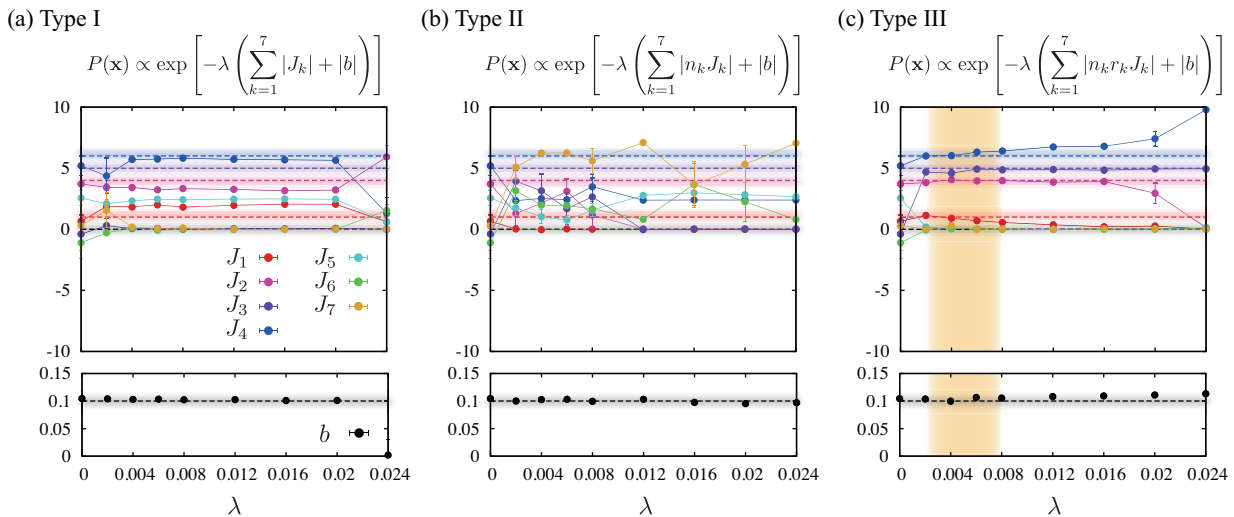} 
\end{center}
\caption{\label{fig:int}
(Color online)
$\lambda$ dependence of estimated spin-spin interactions $J_k$ and $b$ by using the prior distributions of (a) Type I, (b) Type II, and (c) Type III, respectively.
The results are the averages over the four sets of training data.
Dashed lines indicate the input spin-spin interactions $J_k$ and $b$.
The shaded area in (c) indicates that the estimated spin-spin interactions are roughly the same as the values used in the inputted magnetization curve.
}
\end{figure*}

We considered three types of prior distributions for the model parameters $\mathbf{x}$ based on the $l_1$ regularization:
\begin{eqnarray}
&{\rm Type \ I} :& \ P(\mathbf{x}) \propto \exp \left[ - \lambda \left( \sum_{k=1}^7 |J_k| + |b| \right) \right], \label{eq:type1} \\
&{\rm Type \ II} :& \ P(\mathbf{x}) \propto \exp \left[ - \lambda \left( \sum_{k=1}^7 |n_k J_k| + |b| \right) \right], \label{eq:type2} \\
&{\rm Type \ III} :& \ P(\mathbf{x}) \propto \exp \left[ - \lambda \left( \sum_{k=1}^7 |n_k r_k J_k| + |b| \right) \right], \ \ \ \ \ \ \label{eq:type3}
\end{eqnarray}
where $n_k$ is the number of each type of interaction $J_k$ per tetrahedron,
and $r_k$ is the distance between the spins connected by interaction $J_k$.
Table I summarizes the values used in this demonstration.
Type I is the general form of the $l_1$ regularization.
Type II considers the number of interactions per tetrahedron.
In the theoretical model,
the number of interactions in the lattice depends on $J_k$.
Thus, we constructed the $l_1$ regularization from the viewpoint of the sum of the absolute values of the spin-spin interactions (i.e., the number of interactions per tetrahedron was introduced in the prior distribution).
Type III considers both the number and the distance of the interactions.
In general,
the amplitude of the spin-spin interactions decreases as the distance between spins increases.
Therefore,
we assumed that the effect of the long-range interactions on the physical properties was smaller than that of the short-range interactions.
To add a penalty to the long-range interactions,
we introduced the distance between spins in the prior distribution in Type III.

Figure~\ref{fig:int} shows the $\lambda$ dependence of the estimated spin-spin interactions that maximize $P (\mathbf{x} | \{ m^{\rm ex} (H_l) \}_{l \in G_s})$ with $s=1, ..., 4$.
The results are the averages over the spin-spin interactions estimated by the four sets of training data $\mathbf{x}_s^*$ with $s=1, ..., 4$,
and the error bars calculated from the standard deviation.
Large error bars indicate that the estimated spin-spin interactions differ across the four sets of training data.
The dashed lines in Fig.~\ref{fig:int} show the values of the spin-spin interactions used in the inputted magnetization curve,
that is, the solution of the estimation problem.
Between $\lambda=0.002$ and $0.008$ for Type III, which is shown in the shaded area in Fig.~\ref{fig:int} (c),
the estimated spin-spin interactions are roughly the same as the values used in the inputted magnetization curve.
However,
these figures do not conclusively show which estimated spin-spin interactions are plausible for the inputted magnetization curve.

\begin{figure}[]
\begin{center}
\includegraphics[scale=0.8]{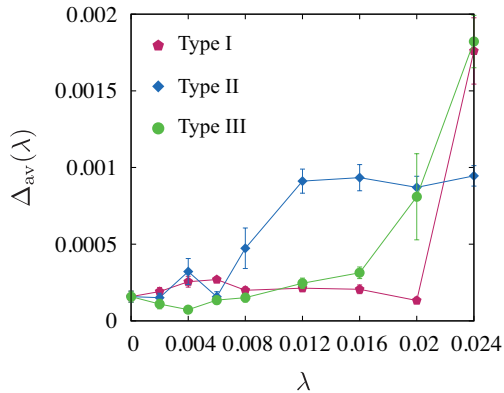} 
\end{center}
\caption{\label{fig:delta}
(Color online)
$\lambda$ dependence of the prediction error $\Delta_{\rm av} (\lambda)$ obtained by cross validation for each type of prior distribution.
$\Delta_{\rm av} (\lambda)$ takes a minimum for $\lambda=0.004$ in Type III.
}
\end{figure}

\begin{figure}[]
\begin{center}
\includegraphics[scale=0.8]{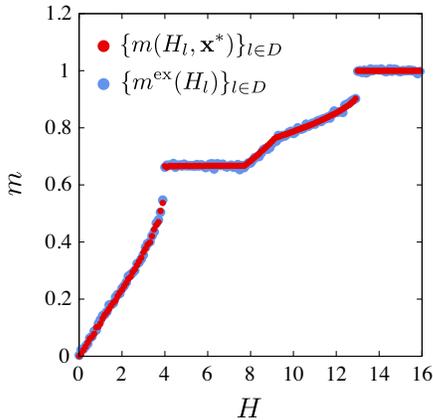} 
\end{center}
\caption{\label{fig:est_mag}
(Color online)
Estimated magnetization curve $\{ m (H_l,\mathbf{x}^*) \}_{l \in D}$ for $L=160$ when $\lambda=0.004$ in Type III and inputted observed magnetization curve $\{ m^{\rm ex} (H_l) \}_{l \in D}$.
Table I shows the estimated spin-spin interactions $\mathbf{x}^*$.
}
\end{figure}

Using cross validation,
we calculated the prediction error $\Delta_{\rm av} (\lambda)$, which is the average of $\Delta^{(s)} (\lambda)$ with $s=1, ..., 4$ defined in Eq.~(\ref{eq:delta_l}).
Figure~\ref{fig:delta} shows the $\lambda$ dependence of $\Delta_{\rm av} (\lambda)$ for each type of prior distribution.
The minimum value of $\Delta_{\rm av} (\lambda)$ occurs at $\lambda=0.004$ in Type III, which produces the highest prediction accuracy.
We conclude that the estimated spin-spin interactions for $\lambda=0.004$ in Type III are the most likely spin-spin interactions for the inputted magnetization curve.
Table I, which shows the estimated spin-spin interactions for $\lambda=0.004$ in Type III where $\Delta^{(s)}(\lambda)$ is the smallest among $s=1, ..., 4$,
confirms that the estimated spin-spin interactions are almost the same as the values used in the inputted magnetization curve.
This demonstrates that our proposed estimation method can estimate the spin-spin interactions with a high accuracy from the observed magnetization curve.
Furthermore,
the estimated values of $J_5$, $J_6$, and $J_7$ are very close to zero,
indicating that the important spin-spin interactions can be selected among the many candidate interactions in the Hamiltonian.
In addition, Fig.~\ref{fig:delta} shows that over-fitting occurs in the case of $\lambda = 0$ when using non-zero values for the additional interaction terms.  
Consequently, over-fitting can be avoided by introducing a regularization term based on the $l_1$ regularization.
Finally, Fig.~\ref{fig:est_mag} compares the estimated magnetization curve for $\lambda=0.004$ in Type III with the inputted magnetization curve.


The model parameters in the Hamiltonian can be estimated with a high accuracy from the observed magnetization curve using the estimation method introduced in Sec. II.
In addition, we confirm that the model parameters are well estimated for certain input magnetization curves with the Hamiltonian defined in Eq.~(\ref{eq:ham}).

\section{Discussion and summary}

We developed a method for estimating the model parameters in the Hamiltonian from the observed magnetization curve.
Bayesian inference is initially used to determine the posterior distribution (i.e., the conditional probability of model parameters in the Hamiltonian for a given observed magnetization curve).
We adopted the $l_1$ regularization as the prior distribution of the model parameters.
In this scheme, the plausible model parameters are determined by maximizing the posterior distribution.
To analyze the posterior distribution,
we adopted the Markov-chain Monte Carlo method and the exchange Monte Carlo method because this combination significantly contributes to finding the global maximum of a posterior distribution where many local maxima exist.
Cross validation is introduced to avoid over-fitting and to determine the hyperparameter in the prior distribution.
Note that the framework of our estimation method can be used not only for magnetization curves but also for any measured data as the input data.

The effectiveness of our method is demonstrated by estimating the spin-spin interactions in the Hamiltonian from the observed magnetization curve.
We used a magnetization curve synthesized by adding Gaussian noise prepared by a theoretical model with fixed spin-spin interactions.
Specifically, we employed a classical Heisenberg model with bilinear and biquadratic interactions on a tetrahedral chain to consider the case where the solution of the estimation problem is known in advance.
Furthermore, 
we introduced three types of prior distributions of the spin-spin interactions based on the $l_1$ regularization.
These prior distributions include the hyperparameter $\lambda$, which controls the strength of the regularization.
The spin-spin interactions are sampled from the posterior distribution for a given $\lambda$ by numerical simulations based on the Monte Carlo method.
We also determined the value of $\lambda$ and the type of the prior
distribution by minimizing the prediction error by cross validation. 
Spin-spin interactions consistent with the input data can be estimated. 
It is noted that our estimation method successfully selects the relevant spin-spin interactions from the many redundant interactions in the Hamiltonian.
The prior distribution taking into account both the number and the distance of the interactions plays a relevant role in this study,
confirming that our method can estimate the spin-spin interactions from an observed magnetization curve with a high accuracy.

In general, it is difficult to obtain physical quantities measured by indirect measurements such as magnetic specific heat and magnetic entropy.
However, our estimation method should predict these quantities once the Hamiltonian for a target magnetic material is obtained. 
Furthermore, 
we can directly obtain the predicted spin-snapshots as well as predict the magnetic structure and structure factor.
Thus, the information from the estimated Hamiltonian should provide preliminary information before using large facilities such as those for neutron diffraction measurements.

Our final objective is to assist with measurements by providing additional estimation information.
However, this research shows only the case of inputting a synthetic magnetization curve obtained from a classical theoretical model, including bilinear and biquadratic interactions.
To achieve our broader objectives, we must address two issues.
First, we need to establish an estimation method that can be used for many types of magnetic materials described by classical or quantum Hamiltonians with many model parameters such as various types of spin-spin interactions, spin-lattice interactions, and spin anisotropy.
In this case, a prior distribution, which estimates the correct model parameters, will depend on the types of target model parameters and the target material.
Thus, it is important to develop a method that provides a useful prior distribution.
Second, we need to improve our proposed estimation method to produce a more robust Hamiltonian that can simultaneously explain various physical properties of a target material.
To achieve this, the input data in the estimation method should include multiple data sets such as some magnetization curves at different temperatures and the magnetic field dependences of the magnetic specific heat.
These are the future directions of our research and will be discussed elsewhere\cite{Tamura-2016}.

Finally, 
we emphasize that our proposed estimation method is a kind of physics-based machine learning technique because it employs a Hamiltonian based on physical laws.
Our proposed method should form the basis for a new applications of machine learning in physics and interdisciplinary physics.
In particular,
in materials informatics\cite{Rajan-2005,Jain-2013}, which is an interdisciplinary field between materials science based on the physical law and informatics,
our proposed estimation method should be a useful tool to establish important concepts.

\section*{Acknowledgement}

We thank Masashi Hase, Haruhiko Kuroe, Hideaki Kitazawa, Noriki Terada, Shu Tanaka, and Masato Okada for the useful discussions.
R. T. was partially supported by a Grant-in-Aid for Scientific Research (C) (Grant No. 25420698). 
K. H. was partially supported by a Grant-in-Aid for Scientific Research
from JSPS, Japan (Grant No. 25120010 and 25610102). 
The computations in the present work were performed on Numerical Materials Simulator at NIMS, 
and the supercomputer at Supercomputer Center, Institute for Solid State Physics, The University of Tokyo.
This work was conducted as part of the ``Materials Research by Information Integration'' Initiative of the Support Program for Starting Up Innovation Hub, Japan Science and Technology Agency.


\end{document}